\documentclass[12pt,psfig]{article}
\usepackage{amsfonts}
\usepackage{amsmath}
\usepackage{makeidx}
\usepackage[dvips]{graphics} 
\usepackage{latexsym,amssymb} 
\usepackage{color} 
\newcommand{\be}{\begin{equation}}
\newcommand{\ee}{\end{equation}}
\newcommand{\bin}[2]{
\left(
\begin{array}{@{}c@{}}
 #1 \\ #2
 \end{array} 
\right)   }

\begin{document}

\vspace*{5mm}

\begin{center} 
{\large \bf Small Fluctuations in $\lambda \phi^{n+1}$ Theory in a Finite Domain:
An Hirota's Method Approach}\\ [10mm]  
\renewcommand{\thefootnote}{\alph{footnote}} 
M.C. Gama$^{1,}$\footnote{e-mail:marceloc@ime.unicamp.br}, J.A. Espich\'an
Carrillo$^{2,}$\footnote{e-mail:jespichan@unac.edu.pe}, A. Maia Jr.$^{1,}$\footnote{e-mail:maia@ime.unicamp.br}, \\ 
[10mm] 
\end{center} 

\begin{center}
\begin{flushleft}
1) \ {\it Instituto de Matem\'atica, Universidade Estadual de Campinas
(UNICAMP) - 13.081-970 - Campinas (SP), Brazil.}\\ 
2) \ {\it Facultad de Ciencias Naturales y Matem\'atica, Universidad Nacional del Callao (UNAC) - Bellavista - Callao, Per\'u.}
\end{flushleft} 
\end{center}

\begin{center} 
{\bf Abstract}
\end{center} 

We present a method to calculate small stationary fluctuations around static solutions describing  bound states 
in a $(1+1)$-dimensional $\lambda \phi^{n+1}$ theory in a finite domain. We also calculate explicitly fluctuations 
for the $\lambda \phi^4$. 
These solutions are written in terms of Jacobi Elliptic functions   
and are obtained from both linear and nonlinear equations. For the linear case we get eingenvalues of a Lam\'e type
Equation and the nonlinear one relies on Hirota's Method.

\baselineskip 0.60cm

\section{Introduction}

\hspace*{0.6cm}It is well known that, for some applications, quantum fields 
can be thought as classical fields upon which are added quantum corrections \cite{jack}.
In addition non-linear field theories are nowadays powerfull tools to describe
fundamental physical theories including cosmology, mainly for the inflationary model
of the universe \cite{lind, kolb}. On the other hand it is also well known that quantum systems 
when placed in finite domains (cavities) can present significant alterations on 
their behaviour. A step further is the study of interacting fields in finite domains ${\cite{carr01}}$.
A modern paradigm is the famous Casimir Effect \cite{casim, casim1}. In 
particular since these corrections are sensitive to boundary conditions an effect 
could expected on bound states of the physical system under consideration.\\
 
In a previous work Carrillo \textit{et al} \cite{nl:a1} obtained a number of solutions of
a $(1+1)$-dimensional $\lambda \phi^4$-Theory in a finite domain in terms of Jacobi 
Elliptic functions. In this work we take a step further in the direction of a 
semi-clasical approach of the theory by studying small fluctuations around one of the 
above mentioned solutions describing its bound states. We think this is enough to exemplify our 
approach to this problem. Nevertheless in this work 
we do not pursue on quantization. Below we show these
fluctuations are described by a non-linear equation. In section 2 we present
the linear case and in section 3 the non-linear case which is solved by Hirota's
method \cite{fl:a2}. In section 4 we present some conclusions and perspectives for
future work.
 
We consider the Lagrangian density ${\cal L}$ of a scalar
field $\phi$ :  
\be
\left.\cal{L}\right.=\frac{1}{2}\partial_{\mu}\partial^{\mu}\phi+\frac{1}{2}M^{2}\phi^{2}-\frac{\lambda}{n+1}\phi^{n+1},
 \label{ft1}
\ee
where $\lambda$ is a coupling constant and $\mu=x,t$. 

From (\ref{ft1}) we derive the following equation of motion, 
\be
-\partial_{\mu}\partial^{\mu}\phi+M^{2}\phi-\lambda\phi^{n}=0.
\label{ft2}
\ee
 With the signature $(+---)$, we can write the equation ($\ref{ft2}$) in dimension $(1+1)$ as:
    
\begin{eqnarray}\label{ft2.1}
	\frac{\partial^{2}\phi}{\partial x^{2}}-\frac{\partial^{2}\phi}{\partial t^{2}}+M^{2}\phi-\lambda\phi^{n}=0.
\end{eqnarray}

The problem now is to solve the equation $(\ref{ft2.1})$ for $\phi$ decomposed in a sum of a static solution 
$\phi_{0}(x)$ and a small fluctuation $\eta(x,t)$. In fact, in order to compare our results in linear case 
with the non-linear one, which will be presented in section $3$ we can, without loss of generality, write
\begin{eqnarray}\label{ft2.3}
	\phi(x,t)=\phi_{0}(x)+\eta(x,t).
\end{eqnarray}
\noindent

Using $(\ref{ft2.3})$ in $(\ref{ft2.1})$ and taking the equations for $\phi_{0}$ and $\eta$ we find:
\begin{eqnarray}\label{ft2.4}
	\frac{d^{2}\phi_{0}}{dx^{2}}+M^{2}\phi_{0}-\lambda\phi_{0}^{n}=0,
\end{eqnarray}
and
\begin{eqnarray}\label{ft2.5}
	\frac{\partial^{2}\eta}{\partial x^{2}}-\frac{\partial^{2}\eta}{\partial t^{2}}+\left(M^{2}-
\lambda n\phi_{0}^{n-1}\right)\eta-\lambda\sum^{n}_{k=2}\bin{n}{k}\phi_{0}^{n-k}\eta^{k}=0.
\end{eqnarray}

Solutions for the equation $(\ref{ft2.4})$ are, in general, given by Abelian Theta Functions using the First Integral 
formalism ${\cite{baker}}$. Now we want to solve the equation $(\ref{ft2.5})$, and, for this we divide the analysis in two 
cases:

\section{The Linear Case} 

\hspace*{0.6cm}Since $\eta$ is a perturbation in the equation $(\ref{ft2.5})$, we assume that $\eta(x,t)=\epsilon \xi(x,t)$ 
where $\epsilon$ is a small constant and then, the last term, which is of order $\left.\cal{O}\right.\left(\eta^{2}\right)$ 
can be neglected. So the equation reduces to:
\begin{eqnarray}\label{ft2.7}
\frac{\partial^{2}\xi}{\partial x^{2}}-\frac{\partial^{2}\xi}{\partial t^{2}}+\left(M^{2}-
\lambda n\phi_{0}^{n-1}\right)\xi=0.
\end{eqnarray}
 
Here we are interested in stationary solutions of form:
\begin{eqnarray}\label{ft2.8}
\xi(x,t)=e^{i \omega t}\psi(x).
\end{eqnarray}

So, Eq.($\ref{ft2.7}$) is reduced to equation for $\psi$:
  
\begin{eqnarray}\label{ft2.9}
	\frac{d^{2}\psi}{dx^{2}}+\left(M^{2}+\omega^{2}-\lambda n \phi_{0}^{n-1}\right)\psi=0.
\end{eqnarray}

Now, we define a parameter $E$ by equation:
\begin{eqnarray}\label{ft2.10}
	\frac{E}{2}=M^{2}+\omega^{2},
\end{eqnarray}
\noindent
and then, the equation $(\ref{ft2.9})$ takes the form:
\begin{eqnarray}\label{ft2.11}
	\frac{d^{2}\psi}{dx^{2}}+\left(\frac{E}{2}-\lambda n \phi_{0}^{n-1}\right)\psi=0.
\end{eqnarray}

This equation can be interpreted as Schr\"odinger Equation for a potencial of the form $\phi_{0}^{n-1}$. The problem now is,
given a static solution $\phi_{0}$ of the equation $(\ref{ft2.4})$, find a general solution of the equation 
$(\ref{ft2.11})$.\\

On the other hand, $\lambda\phi^4$-Theory $(n=3)$, the general static solutions of the classical equation of motion to 
the field $\phi$ are given by ${\rm sn}$-type elliptic functions \cite{fl:a1}
\be
\phi_{0}(x) = \pm\,\frac{M\,\sqrt{2c}}{\sqrt{\lambda}\sqrt{1+\sqrt{1-2c}}}
\,{\rm sn}\,\left( \frac{M\,x}{\sqrt{2}}\,\sqrt{1+\sqrt{1-2c}},l\right),
\label{ft3} 
\ee
where $c$ is a parameter belonging to interval $(0,\frac{1}{2}]$ and
\be
l = \frac{1}{-1+\frac{1+\sqrt{1-2c}}{c}}.
\label{ft4}
\ee

From this relation, clearly $l \in (0,1]$.\\

Now, substituting $\phi_{0}(x)$ given by Eq. $(\ref{ft3})$ in Eq. $(\ref{ft2.11})$ with $n=3$ we get:
\be
\frac{d^{2}\,\psi(\alpha)}{d\alpha^{2}} = \left( 6\,l\,{\rm sn}^{2}(\alpha,l) -
\frac{(1+l)}{2}\,E \right)\,\psi(\alpha),   
\label{ft8} 
\ee
where we have used also Eq. ($\ref{ft2.10}$) and the change of
variable 
\begin{eqnarray*}
\alpha = \frac{M\,x}{\sqrt{1+l}}.
\end{eqnarray*}

In the literature the general form of Eq. (\ref{ft8}) is the Lam\'e differential 
equation, which is given by ${\cite{fl:aa}}$ 
\be
\frac{d^{2}\,\Lambda(\alpha)}{d\alpha^{2}} - \left( m(m+1)\,k^2\,{\rm
sn}^{2}(\alpha,k) + C \right)\,\Lambda(\alpha)=0,   
\label{ft9} 
\ee
where $m$ is a positive real number, $k^2$ is the parameter of the Jacobian
Elliptic Function ${\rm sn}$, and $C$ is an arbitrary constant. Therefore,
comparing (\ref{ft8}) with (\ref{ft9}) we obtain, $m =2$ and $C =
-\,\frac{(1+l)}{2}\,E$.\\

The solutions of the Lam\'e equation have been studied in the literature \cite{fl:a3}. 
With a few algebraic manipulations we get the following eingenfunctions and their 
respectives eingenvalues:
\begin{eqnarray*}
(1) \ \ \eta_l(x,t) = \epsilon\exp( i\sqrt{\frac{3}{1+l}}Mt){\rm sn}
(\frac{Mx}{\sqrt{1+l}},l){\rm cn}(\frac{Mx}{\sqrt{1+l}},l)
\end{eqnarray*}
\be
 \ \ \ \ \mbox{for} 
\ \ \ \ \omega_{1}^{2} = \left( \frac{3}{1+l}\right)M^2,
\label{ft10}
\ee

\begin{eqnarray*}
(2) \ \ \eta_l(x,t) =\epsilon  \exp( i\sqrt{\frac{3l}{1+l}}Mt)
{\rm sn}(\frac{M\,x}{\sqrt{1+l}},l)\,{\rm dn}
(\frac{M\,x}{\sqrt{1+l}},l)
\end{eqnarray*}
\be
 \ \ \ \ \ \ \mbox{for} \ \ \ \ \ \ \omega_{2}^{2} = 
\left( \frac{3\,l}{1+l}\right)\,M^2,
\label{ft11}  
\ee

\begin{eqnarray}
(3) \ \ \eta_l(x,t) = \epsilon {\rm cn}(\frac{M\,x}{\sqrt{1+l}},l)\,{\rm dn}
(\frac{M\,x}{\sqrt{1+l}},l) \ \ \ \ \ \ \mbox{for} \ \ \ \ \ \
\omega_{3}^{2} = 0,
\label{ft12}
\end{eqnarray}

\begin{eqnarray*}
(4) \ \ \eta_l(x,t) =\epsilon \exp(iw_{4}t)
\left({\rm sn}^{2}(\frac{Mx}{\sqrt{1+l}},l) - \frac{1+l+\sqrt{l^2 -l+1}}{3l}\right) 
\end{eqnarray*}
\be
\ \ \ \ \ \ \mbox{for} \ \ \ \ \ \ \omega_{4}^{2} = \left( \frac{1+l-2\sqrt{l^2 -
l+1}}{1+l}\right)\,M^2, 
\label{ft13}
\ee

\begin{eqnarray*}
(5) \ \ \eta_l(x,t) = \epsilon \exp(i\omega_{5}t)
\left({\rm sn}^{2}(\frac{Mx}{\sqrt{1+l}},l) - \frac{1+l-\sqrt{l^2 -l+1}}{3l}\right) 
\end{eqnarray*}
\be
\ \ \ \ \ \ \mbox{for} \ \ \ \ \ \ \omega_{5}^{2} = \left( \frac{1+l+2\sqrt{l^2 -
l+1}}{1+l}\right)\,M^2. 
\label{ft14}
\ee

From the previous relations, it is worth to note  that taking $l=1$ in (\ref{ft12}),
(\ref{ft13}) and (\ref{ft10}), (\ref{ft11}), we recover respectively, the 
eigenfunctions and energy levels of the ground-state and the first excited  state 
(static case) of the Dashen-Hasslacher-Neveu (DHN model)${\cite{nl:dhn}}$. For 
the fifth case, we observe that for $l=1$, $\omega_{5}^{2} = 2M^2$, which corresponding to 
the continuum part of the espectrum.

\section{Bound States Fluctuations by Hirota's Method}

\hspace*{0.6cm} Now let us come back to the nonlinear model of the Equation $(\ref{ft2.5})$. Again we will find the 
fluctuations $\eta$, but now using a slight modification of the Hirota's Method \cite{fl:a2}. This is done as follows. 
Firstly we make a dependent variable transformation, namely,
\be
	\eta(x,t)=\frac{\Phi_{x}(x,t)}{\Phi^{n-2}(x,t)}.
\label{ft15}
\ee

In original Hirota's Method, a $\beta$ parameter enters the above equation as a multiplicative constant in order 
to simplify the differential equations obtained. In our method, this is not necessary, so the $\beta$ parameter does 
not appear in the equation $(\ref{ft15})$.\\
 
Thus, substituting (\ref{ft15}) in (\ref{ft2.5}), we get the following equation  
\begin{eqnarray*}
	\Phi^{2}\Phi_{xxx}+(6-3n)\Phi\Phi_{x}\Phi_{xx}+(n-1)(n-2)\Phi_{x}^{3}-(n-1)(n-2)\Phi_{x}\Phi_{t}^{2}+
\end{eqnarray*}
\begin{eqnarray*}	
	+(n-2)\Phi\Phi_{x}\Phi_{tt}+(2n-4)\Phi\Phi_{t}\Phi_{xt}-\Phi^{2}\Phi_{xtt}+M^{2}\Phi^{2}\Phi_{x}+
\end{eqnarray*}
\begin{eqnarray}\label{ft15.1}
-\lambda\Phi^{n}\sum^{n}_{k=1}\bin{n}{k}\phi_{0}^{n-k}\Phi_{x}^{k}\Phi^{2k-nk}=0.
\end{eqnarray}

The Hirota's Method prescribes that $\Phi(x)$ can be expanded as a power series:
\begin{eqnarray}
\Phi(x,t) = 1 + \gamma\,f_1 + \gamma^2\,f_2 + ... \quad,
\label{ft17}
\end{eqnarray}
where $f_i(x,t)$ are unknown functions $(i = 1,2,3, ...)$ and $\gamma$ is an 
arbitrary constant ${\cite{fl:a2}}$. Substituting $(\ref{ft17})$ into $(\ref{ft15.1})$ and equating 
the coefficients for every degree of $\gamma$ to zero, we obtain an infinite 
system of differential equations (we show below only the first two):
\be \label{ft17.1}
f_{1,xxx}-f_{1,xtt}+\left(M^{2}-\lambda n\phi_{0}^{n-1}\right)f_{1,x}=0,
\ee
and
\begin{eqnarray*} 
f_{2,xxx}-f_{2,xtt}+\left(M^{2}-\lambda n\phi_{0}^{n-1}\right)f_{2,x} &=& 
\end{eqnarray*}
\begin{eqnarray*}
& = & 2f_{1}f_{1,xtt}-2\left(M^{2}-\lambda n\phi_{0}^{n-1}\right)f_{1}f_{1,x}+
\end{eqnarray*}
\begin{eqnarray*}
+\left((3n-6)f_{1,xx}-(n-2)f_{1,tt}\right)f_{1,x}-(2n-4)f_{1,t}f_{1,xt}-2f_{1,xxx}f_{1}
\end{eqnarray*}
\begin{eqnarray}
-\lambda\frac{n(n-1)}{2}\phi_{0}^{n-1}f_{1,x}^{2}.
\end{eqnarray}
For now, we especialize for $\lambda\phi^4$ theory. In this case the above equation reduces to:
\begin{eqnarray*}
-\Phi_{xtt}\Phi^{2} + 2\Phi\Phi_{xt}\Phi_{t} + \Phi\Phi_{x}\Phi_{tt} -
2\Phi_{x}\Phi^{2}_{t} \ +  
\end{eqnarray*}
\begin{eqnarray}  
\Phi_{xxx}\Phi^2 - 3\,\Phi\Phi_{x}\Phi_{xx} + (2 - \lambda)
\Phi_{x}^{3} + M^{2}\Phi_{x}\Phi^{2} - 3\lambda\phi_{0}^{2}\Phi_{x}\Phi^{2} - 
3\lambda\phi_{0}\Phi_{x}^{2}\Phi = 0.  
\label{ft16}  
\end{eqnarray}
 
The equations for $f_{1}$ and $f_{2}$ are now: 
\be 
- f_{1,xtt} + f_{1,xxx} + (M^{2} -
3\,\lambda\,\phi^{2}_{0})\,f_{1,x} = 0, 
\label{ft18}
\ee
\begin{eqnarray*} 
f_{2,xxx}-f_{2,xtt}+\left(M^{2}-3\lambda\phi_{0}^{2}\right)f_{2,x}= 2f_{1}f_{1,xtt}-
2\left(M^{2}-3\lambda \phi_{0}^{2}\right)f_{1}f_{1,x}+
\end{eqnarray*}
\begin{eqnarray}
+\left(3f_{1,xx}-f_{1,tt}\right)f_{1,x}-2f_{1,t}f_{1,xt}-2f_{1,xxx}f_{1}-3\lambda\phi_{0}^{2}f_{1,x}^{2}
\end{eqnarray}\label{ft19}
and so on for higher orders $\left.\cal{O}\right.(\gamma^{3})$.\\

Notice that the function $f_i$ will be determined by the previous functions $\left\{f_{1},f_{2},...f_{i-1}\right\}$ only. 
So, we have a infinite set of recurrent equations. The next step is the usual one, as the first part of this work 
{linear case). Thus, defining the new function
\be
\xi(x,t) = f_{1,x},
\label{ft20}
\ee
the Eq. (\ref{ft18}) can be written as
\be
- \xi_{tt} + \xi_{xx} + (M^{2} -
3\,\lambda\,\phi^{2}_{0})\,\xi = 0. 
\label{ft21}
\ee

Observe that the above equation is pretty the same as
linearized equation $(\ref{ft2.7})$. Since $\xi$ satisfies $(\ref{ft2.7})$ we can then use 
the solutions $((\ref{ft10}) - (\ref{ft14}))$ to obtain the functions $f_1$, which 
are given by: 
\begin{eqnarray*}
\hspace*{1.4cm}(1) \ \ f_1(x,t) = - \frac{\sqrt{1+l}}{M\,l}\,
\exp( i\sqrt{\frac{3}{1+l}}\,M\,t)\,
{\rm dn}(\frac{M\,x}{\sqrt{1+l}},l),    
\end{eqnarray*}
\be
\ \ \ \ \ \ \mbox{for} \ \ \ \ \ \ w_{1}^{2} = \left( 
\frac{3}{1+l}\right)\,M^2.  
\label{ft24}
\ee

\begin{eqnarray*}
\hspace*{1.2cm}(2) \ \ f_1(x,t) = - \frac{\sqrt{1+l}}{M}\,
\exp( i\sqrt{\frac{3\,l}{1+l}}\,M\,t)\,
{\rm cn}(\frac{M\,x}{\sqrt{1+l}},l),
\end{eqnarray*}
\be
\ \ \ \ \ \ \mbox{for} \ \ \ \ \ \  w_{2}^{2} = \left(
\frac{3\,l}{1+l}\right)\,M^2. 
\label{ft26}
\ee

\begin{eqnarray}
(3) \ \ f_1(x) = \frac{\sqrt{1+l}}{M}\,{\rm sn}(\frac{M\,x}{\sqrt{1+l}},l) 
\ \ \ \ \ \ \mbox{for} \ \ \ \ \ \ w_{3}^{2} = 0.
\label{ft27}
\end{eqnarray}

\begin{eqnarray*}
(4) \ \ f_1(x,t) = \exp( i\sqrt{\frac{1+l-2\sqrt{l^2 -l+1}}{1+l}}Mt)
\biggl( \frac{\sqrt{1+l}}{l\,M}( \frac{Mx}{\sqrt{1+l}} - \biggr.
\end{eqnarray*}
\begin{eqnarray*}
\biggl. E( am \frac{Mx}{\sqrt{1+l}},l))  - 
 \frac{(1+l+\sqrt{l^2 -l+1})\,x}{3l} \biggr),
\end{eqnarray*}
\be
\mbox{for} \ \ \ \ \ \ w_{4}^{2} = \left( 
\frac{1+l-2\sqrt{l^2 - l+1}}{1+l}\right)\,M^2.
\label{ft29}
\ee

\begin{eqnarray*}
(5) \ \ f_1(x,t) = \exp( i\sqrt{\frac{1+l+2\sqrt{l^2 -l+1}}{1+l}}Mt)
\biggl( \frac{\sqrt{1+l}}{l\,M}( \frac{Mx}{\sqrt{1+l}} - \biggr.
\end{eqnarray*}
\begin{eqnarray*}
\biggl. E( am \frac{Mx}{\sqrt{1+l}},l))-  
\frac{(1+l-\sqrt{l^2 -l+1})\,x}{3l} \biggr),
\end{eqnarray*}
\be
\mbox{for} \ \ \ \ \ \ w_{5}^{2} = \left( 
\frac{1+l+2\sqrt{l^2 - l+1}}{1+l}\right)\,M^2,
\label{ft31}
\ee
where, in all cases the constants of integration are taken as equal to zero.

The next step consists in the determination of the function $f_2$. In this 
case, using the Eq. (\ref{ft18}) into Eq. (\ref{ft19}) results the equation for $f_{2}$: 
\begin{eqnarray} 
f_{2,xxx}-f_{2,xtt}+\left(M^{2}-3\lambda\phi_{0}^{2}\right)f_{2,x}= 
\left(3f_{1,xx}-f_{1,tt}\right)f_{1,x}-2f_{1,t}f_{1,xt}-3\lambda\phi_{0}^{2}f_{1,x}^{2}
\end{eqnarray}

One can see that substituting the functions $f_1$ given by (\ref{ft24}) - (\ref{ft31})
in above equation, we obtain inonhomogeneous differential equations for the 
function $f_2$. Unfortunately due to the its great complexity, it was not 
possible to obtain the contribution of $f_2$ with this approach.
Nevertheless we have obtained solutions for the nonlinear equation (\ref{ft2.5}),
considering only the contribution of the function $f_1$. On the other hand, $\gamma$ 
is an arbitrary constant in the Hirota's  Method, since the series in equation $(\ref{ft17})$ 
is just a formal series ${\cite{fl:a2}}$. On the other hand, as mentioned above, in Eq. $(\ref{ft2.3})$, 
the field $\eta(x,t)$  is a small fluctuation on a static solution $\phi_{0}(x)$. 
We have found solutions below whose amplitudes are of order $\gamma$ (first approximation). 
So, for our application, $\gamma$ is a small number controlling the fluctuation amplitude. Using (\ref{ft15}), 
we obtain the corresponding fluctuations $\eta$, which are given by:
\begin{eqnarray}
(1) \ \ \eta_{l}(x,t) = 
\frac{\gamma \exp( i\sqrt{\frac{3}{1+l}}\,M\,t)\,
{\rm sn}(\frac{Mx}{\sqrt{1+l}},l)\,{\rm cn}(\frac{Mx}{\sqrt{1+l}},l)}
{1-\gamma\frac{\sqrt{1+l}}{M\,l}\,\exp( i\sqrt{\frac{3}{1+l}}\,M\,t)\,
{\rm dn}(\frac{Mx}{\sqrt{1+l}},l)}, 
\label{ft32}
\end{eqnarray}

\begin{eqnarray}
(2) \ \ \eta_{l}(x,t) = 
\frac{\gamma\exp( i\sqrt{\frac{3\,l}{1+l}}\,M\,t)\,
{\rm sn}(\frac{Mx}{\sqrt{1+l}},l)\,{\rm dn}(\frac{Mx}{\sqrt{1+l}},l)}
{1-\gamma\frac{\sqrt{1+l}}{M}\,\exp( i\sqrt{\frac{3\,l}{1+l}}\,M\,t)\,
{\rm cn}(\frac{M\,x}{\sqrt{1+l}},l)},
\label{ft33}
\end{eqnarray}

\begin{eqnarray}
(3) \ \ \eta_{l}(x) = 
\frac{\gamma {\rm cn}(\frac{Mx}{\sqrt{1+l}},l)\,{\rm dn}
(\frac{Mx}{\sqrt{1+l}},l)}{1-\gamma\frac{\sqrt{1+l}}{M}\,{\rm sn}
(\frac{Mx}{\sqrt{1+l}},l)},
\label{ft34}
\end{eqnarray}

\begin{eqnarray}
(4) \ \eta_{l}(x,t) = \frac{
 \gamma\exp( i\sqrt{\frac{1+l-2\sqrt{l^2 -l+1}}{1+l}}Mt)
\left( {\rm sn}^{2}(\frac{Mx}{\sqrt{1+l}},l) - \frac{1+l+\sqrt{l^2 -l+1}}{3l}\right)}
{1 +\gamma \exp( i\sqrt{\frac{1+l-2\sqrt{l^2 -l+1}}{1+l}}Mt)
\biggl( \frac{x(2-l-\sqrt{l^2 -l+1})}{3l}
-\frac{\sqrt{1+l}}{Ml}E( am \frac{Mx}{\sqrt{1+l}},l)\biggr)},
\label{ft35}  
\end{eqnarray}

\begin{eqnarray}
(5) \ \eta_{l}(x,t) = \frac{
\gamma\exp( i\sqrt{\frac{1+l+2\sqrt{l^2 -l+1}}{1+l}}Mt)
\left( {\rm sn}^{2}(\frac{Mx}{\sqrt{1+l}},l) - 
\frac{1+l-\sqrt{l^2 -l+1}}{3l}\right)} {1 + \gamma\exp( i\sqrt{\frac{1+l+2\sqrt{l^2
-l+1}}{1+l}}Mt) \biggl( \frac{x(2-l+\sqrt{l^2 -l+1})}{3l}
-\frac{\sqrt{1+l}}{Ml}E( am \frac{Mx}{\sqrt{1+l}},l)\biggr)}.
\label{ft36}  
\end{eqnarray}

 Now observe that, at first aproximation $(0<\gamma<<1)$, the correspondent fluctuations of linear and non-linear cases coincides, which implies $\gamma=\epsilon$ in  equations $(\ref{ft32}-\ref{ft36})$. This show that linear case is, in fact, an first order approximation of the non-linear one.\\
  
On the other hand, in \cite{nl:a1} the authors showed that imposing Dirichlet
boundary conditions on the field $\phi$, given by (\ref{ft3}), confined to a box of 
size $L$, it must satisfy the condition
\be
ML = 4\sqrt{1+l}K(l),
\label{con}
\ee
where $4K(l)$ is a period of the Jacobi Elliptic Function ${\rm sn}(u,l)$ \cite{abra}.
So, imposing the same boundary conditions for the fluctuations $\eta$ (same kind
of confinement), implies that the same $l = l(L)$, obtained from above 
relation, must be used for $\eta$. It is not difficult to see  from our calculation, 
that, when $\eta$ is
confined, the only consistent  linear and nonlinear fluctuations satisfying the  
condition (\ref{con}) are those given by (\ref{ft10}), (\ref{ft11}), (\ref{ft32}) and 
(\ref{ft33}), for they have the Jacobi Elliptic function ${\rm sn}(u,l)$, from which the relation (\ref{con}) 
comes from.\\

In the more general case, of the $\lambda\phi^{n+1}$ theory, the first approximation for fluctuations can be 
calculated from Eq.$(\ref{ft15})$ and Eq.$(\ref{ft17})$. It is given by
  
\be
\eta(x,t)= \gamma\frac{ f_{1,x}}{(1+\gamma f_{1})^{n-2}}.
\label{ft37}
\ee  
\noindent
where the functions $f_{1}$, obtained by Hirota's Method.   
       
\section{Conclusions}

\hspace*{0.6cm}We developed an adaptation of the Hirota's Method in order to calculate small fluctuations 
around static bound states in a $\lambda \phi^{n+1}$-theory. For the case $n=3$, we calculate those fluctuations in a 
Finite Domain (interval) in (1+1)-dimensions in order to solve a non-linear equation of motion. We showed 
that those fluctuations are written in terms of Jacobi Elliptic functions. Imposing that fluctuations must 
satisfy the same boundary conditions of the unperturbed bound states we can select a subset of solutions. 
We must stress that, in the limit $l=1$ ones get the results by Knyazev on fluctuations around a kink solution \cite{fl:a4}, for
the fluctuations in the Eqs. (\ref{ft32}) and (\ref{ft34}). Even in this case (kink) we have obtained three new 
fluctuations, (\ref{ft33}), (\ref{ft35}) and (\ref{ft36}). Dirichlet conditions imply the only possibility for fluctuation 
(\ref{ft35}) is $l=1$, so it does not exist in finite domains. The solution (\ref{ft36}) there exists 
in any domain.
 
\vspace*{0.5cm}
\noindent{\Large \bf Acknowledgments}

This work was partially supported by CAPES (Coordena\c{c}\~ao de Aperfei\c{c}oamento de Pessoal de N\'{\i}vel Superior).

\end{document}